\documentclass[usenatbib]{mn2e}
\usepackage{mathrsfs,amsmath,amssymb,amstext}
\usepackage{lscape}
\usepackage{natbib}
\usepackage{color}
\usepackage{float}
\usepackage{graphicx}
\usepackage{gensymb}
\usepackage[justification=centering]{caption}
\usepackage[flushleft]{threeparttable}
\usepackage[breaklinks,colorlinks,citecolor=blue,linkcolor=magenta]{hyperref} 
\usepackage{footnote}
\usepackage{movie15}
\usepackage{animate}
\makesavenoteenv{tabular}

\usepackage[all]{hypcap}

\begin{document}

\title[Double X/P Structures]{Double X/Peanut Structures in Barred Galaxies -- Insights from an $N$--body Simulation}
\author[Ciambur et al.]{Bogdan C. ~Ciambur$^{1,2}$\thanks{E-mail: bogdan.ciambur@obspm.fr}, Francesca ~Fragkoudi$^{3}$, Sergey ~Khoperskov$^{4}$,
 \newauthor
Paola ~Di Matteo$^{2}$, Fran\c coise ~Combes$^{1,5}$ \\ \\
%LERMA, Observatoire de Paris, CNRS, PSL University, Sorbonne University,  Paris, France, 75014 Paris, France \\
$1$ LERMA, Observatoire de Paris, CNRS, PSL University, Sorbonne Universit\'es, UPMC Univ. Paris 06, F-75014, Paris, France\\
$2$ GEPI, Observatoire de Paris, PSL Research University, Place Jules Janssen, F-92190, Meudon, France\\
$3$ Max-Planck-Institut f\"ur Astrophysik, Karl-Schwarzschild-Str. 1, D-85748, Garching, Germany\\
$4$ Max Planck Institute for Extraterrestrial Physics, D-85741, Garching, Germany\\
$5$ Coll\`ege de France, 12 Place Marcelin-Berthelot, F-75005 Paris, France}
\maketitle

\begin{abstract}

Boxy, peanut-- or X--shaped ``bulges'' are observed in a large fraction of barred galaxies viewed in, or close to, edge-on projection, as well as in the Milky Way. They are the product of dynamical instabilities occurring in stellar bars, which cause the latter to buckle and thicken vertically. Recent studies have found nearby galaxies that harbour two such features arising at different radial scales, in a nested configuration. In this paper we explore the formation of such double peanuts, using a collisionless $N$--body simulation of a pure disc evolving in isolation within a live dark matter halo, which we analyse in a completely analogous way to observations of real galaxies. In the simulation we find a stable double configuration consisting of two X/peanut structures associated to the same galactic bar -- rotating with the same pattern speed -- but with different morphology, formation time, and evolution. The inner, conventional peanut-shaped structure forms early via the buckling of the bar, and experiences little evolution once it stabilises. This feature is consistent in terms of size, strength and morphology, with peanut structures observed in nearby galaxies. The outer structure, however, displays a strong X, or ``bow-tie'', morphology. It forms just after the inner peanut, and gradually extends in time (within 1 to 1.5 Gyr) to almost the end of the bar, a radial scale where ansae occur. We conclude that, although both structures form, and are dynamically coupled to, the same bar, they are supported by inherently different mechanisms.

\end{abstract}

\begin{keywords}

{galaxies: structure -- galaxies: bulges -- galaxies: evolution -- galaxies: kinematics and dynamics}

\end{keywords}

\newcommand{\elli}{$\textsc{Ellipse}$}
\newcommand{\iso}{$\textsc{Isophote}$}
\newcommand{\bmo}{$\textsc{Bmodel}$}
\newcommand{\cmo}{$\textsc{Cmodel}$}
\newcommand{\ifit}{$\textsc{Isofit}$}
\newcommand{\rf}{\textcolor{red}{reference}}
\newcommand{\rad}{$R_{\rm maj}$}
\newcommand{\Rpiint}{$R_{{\it \Pi},{\rm int}}$}
\newcommand{\Msun}{$M_{\odot}$}

\section{Introduction}\label{sec:introduction}
% \begin{itemize}
 
% \item Discuss some simulation works/explanation of phenomenon (\citealt{Combes&Sanders1981}, \citealt{Combes+1990}), mention observational evidence for bar buckling \cite{Erwin&Debattista2016}.
 %\item Recent systematic studies of host XP galaxy properties and evolution (\citealt{Erwin&Debattista2017}, \citealt{Kruk+2019}) - some general conclusions?.
 %\item Explain observed double peanut structures in \cite{Ciambur&Graham2016}, hereafter CG16.
 %\item Recently, a double peanut configuration has been detected in the face-on galaxy NGC 1291 \citep{Mendez-Abreu+2019}, via its face-on signature \citep{Debattista+2005}.
 %\item Explain how we will use a pure disc simulation to investigate this particular -- double peanut -- stellar configuration.
% \item Summarise the main sections 
 %\end{itemize}
 
 %\vspace{20cm}

Barred galaxies often display boxy, X-- or peanut--shaped ``bulges'' when observed in an edge-on orientation of their stellar disc. Unlike conventional, axisymmetric galaxy bulges -- either built secularly from the disc (typically oblate and rotating ``pseudobulges''), or constituting the remnants of past mergers (spheroidal classical bulges) -- X/peanut (X/P) structures are believed to be the vertically thickened and non-axisymmetric inner parts of bars. Early studies with $N$--body simulations have shown that peanuts arise naturally from bars as the latter experience vertical (i.e., perpendicular to the disc plane) resonances, particularly the vertical 2:1 inner Lindblad resonance (vILR) (\citealt{Combes&Sanders1981}, \citealt{Combes+1990}) or buckling instabilities (\citealt{RahaEA1991}, \citealt{MerrittSellwood1994}, \citealt{Martinez-Valpuesta+2007}), and can have a significant effect on their host galaxies (\citealt{Fragkoudi+2015}, \citealt{Fragkoudi+2016}). In parallel, the (bar--peanut) connection has been extensively supported observationally (\citealt{Shaw1987}, \citealt{DettmarBarteldress1990}, \citealt{KuijkenMerrifield1995}, \citealt{BureauFreeman1999}, \citealt{BureauAthanassoula2005}), with \cite{Erwin&Debattista2016} recently showing observations of galaxy bars in the act of buckling to form peanut structures. The field has advanced sufficiently in recent times to perform systematic studies on the X/P host galaxy population. \cite{Erwin&Debattista2017} have shown that the incidence of X/P structures in barred galaxies is a strong function of their host's stellar mass and Hubble type (the fraction being higher in more massive, and earlier--type, disc galaxies), while more recently the frequency of occurrence of X/P structures has been probed as a function of redshift, from observations ranging from the local Universe out to $z \sim 1$ (\citealt{Kruk+2019}), as well as in cosmological simulations \citep{Fragkoudi+2019b}. As a rather typical spiral barred galaxy, the Milky Way too displays a peanut--shaped ``bulge'' (\citealt{Okuda+1977}, \citealt{Dwek+1995}, \citealt{Binney+1997}, \citealt{Babusiaux&Gilmore2005}, \citealt{Lopez-Corredoira+2005}, \citealt{Gerhard2015}, \citealt{Gonzalez&Gadotti2016}, \citealt{Ness&Lang2016}), and our privileged placement within its disk has allowed us to study its bulge in exquisite detail, in terms of its large-scale morphology in the near-infrared (\citealt{CiamburGraham&Bland-Hawthorn2017}), or as inferred through tracers (\citealt{Mcwilliam&Zoccali2010}, \citealt{Ness+2012}, \citealt{Lopez-Corredoira+2005}, \citealt{WeggGerhard&Portail2015}). Stellar population studies have shown that the vast majority of stars in the Galactic bulge originate from the disc (\citealt{Shen+2010}, \citealt{Ness+2013}, \citealt{DiMatteo+2014},  \citealt{DiMatteo2016}, \citealt{Fragkoudi+2017c}, \citealt{Fragkoudi+2018}), further support for its formation through a secular channel.

X/P structures leave a measurable imprint in the light distribution of galaxies, specifically in the $6^{\rm th}$ order Fourier harmonic (the cosine term coefficient, $B_6$) of the one-dimensional intensity distribution along an isophote. Exploiting this, \cite{Ciambur&Graham2016}, hereafter CG16, have introduced a framework to extract quantitative parameters of X/P structures, such as size and strength, directly from galaxy images. Applying this method on a sample of twelve nearby galaxies previously known in the literature to host peanut--shaped bulges, they showed that the X/P parameters are correlated and broadly define structural scaling relations. Interestingly, two galaxies in this sample, NGC~128 and NGC~2549, were found to host a double, nested peanut configuration, i.e., a small scale X/P bulge embedded within a larger one. While only recently simulations of barred galaxies are seen to develop double X--shaped features under certain circumstances (e.g., \citealt{Debattista+2017}, \citealt{Smirnov&Sotnikova2018}, \citealt{Parul+2020}), the face-value interpretation of CG16 at the time was that each of the two co-existing X/P structures is associated with a distinct bar component, as double bar galaxies had been known for long to occur (\citealt{Friedli+1996}, \citealt{Erwin+2001}, \citealt{Erwin2011}). Indeed, in a recent paper \cite{Mendez-Abreu+2019} have detected a double peanut configuration in the face--on galaxy NGC~1291 (using the $face$--$on$ signature of peanut galaxies, as introduced in \citealt{Debattista+2005}), which indeed hosts a large--scale, and a second embedded nuclear, bar. Nevertheless, the scale ratio of the two peanuts in NGC~128 and especially in NGC~2549 ($\approx 3:1$) is rather low for a (large scale+nuclear) bar configuration, and thus warrants further investigation. 

In this paper we make use of a collisionless $N$--body simulation of a stellar disc evolving in isolation within a live dark matter halo, to study the development and stellar properties of a stable and persistent double (nested) configuration of X/peanut--shaped structures. In Section \ref{sec:data} we detail the simulation setup as well as the various secular processes it undergoes while evolving in isolation. In Section \ref{sec:analysis} we analyse mock-images generated from the simulation at an epoch where both nested peanuts have formed and stabilised, using a framework equivalently applicable to observational data. Here we model the radial surface density profile, and extract the parameters (size and strength) of the two nested X/P structures. In Section \ref{sec:discussion} we interpret and discuss the results, investigating the dynamics and formation of this nested configuration, and quantitatively comparing the X/P parameters with those of the CG16 sample. Section \ref{sec:conclusions} concludes this study.

\section{Simulation}\label{sec:data}

We explore a purely collisionless $N$-body simulation of a multicomponent model for a galaxy consisting of three co-spatial disk populations: cold, warm, and hot disks represented by $10 \times 10^6$, $6 \times 10^6$, $4 \times 10^6$ particles respectively. Such a setup is convenient for our analysis because each kinematic component has a different response to dynamical instabilities such as the vertical buckling instability of interest in this work (see \citealt{Fragkoudi+2017b}). As such, the X/P development can be followed progressively and disentangled from other processes, as we shall detail further in the following Section. Further, a multiple-component disc additionally provides a more realistic configuration (e.g., \citealt{DiMatteo+2019}), and reproduces observations \cite{Fragkoudi+2018}, thus rendering comparison with real galaxies all the more pertinent.

%\textbullet As each has a different response/sensitivity to dynamical instabilities, like the X/P instability of interest here, convenient to study progressively these secular processes, as we will develop in the following section.
%\textbullet In addition, potentially more realistic configuration, similar to observed galaxies and  Milky Way (Paola's works)

\subsection{Simulation Set-up}

Initially stellar particles are redistributed following a Miyamoto-Nagai density profile~\citep{MiyamotoNagai1975} that has a characteristic scale length of $h_{\rm ini} = $ $4.7$, $2.3$, and $2.3$ kpc, vertical thicknesses of $0.3$, $0.6$ and $0.9$~kpc and masses of $2.5$, $1.5,$ and $1.0 \times 10^{10}$~\Msun, respectively.  Our simulation also includes a live dark matter halo~($5\times 10^6$ particles) whose density distribution follows a Plummer sphere~\citep{1911MNRAS..71..460P}, with a total mass of $1.6\times 10^{11}$~\Msun and a characteristic radius of $10$~kpc. The stellar particles have a mass of $2.5\times10^3$~\Msun, while dark matter particles have a mass of $3.2\times10^4$~\Msun. For the $N$-body system integration, we used our parallel version of the TREE code with multithread usage under the AVX instructions. In the simulation we adopted the standard opening angle $\theta=0.5$ and a gravitational softening parameter equal to $10$~pc. For the time integration, we used a leapfrog integrator with a fixed step size of $0.1$~Myr.

\subsection{Evolution in Isolation}\label{sec:evoln}

%\begin{itemize}
%\item N-body simulation
%\item Dark matter halo component (Plummer) potential
%\item Stellar component: pure disc: Miyamoto-Nagai \citep{MiyamotoNagai1975} initial density profile
%\item Pure disc -- why: avoid complications/degeneracies in structural decomposition, which may arise from the presence of a classical bulge component
%\item Evolves in isolation: each structure which appears at later times is solely due to secular evolution
%\item Disc has 3 kinematic components: thin, intermediate and thick. Scale lengths and scale heights [!]

\begin{figure*}
	\centering
	\includegraphics[width=0.645\columnwidth]{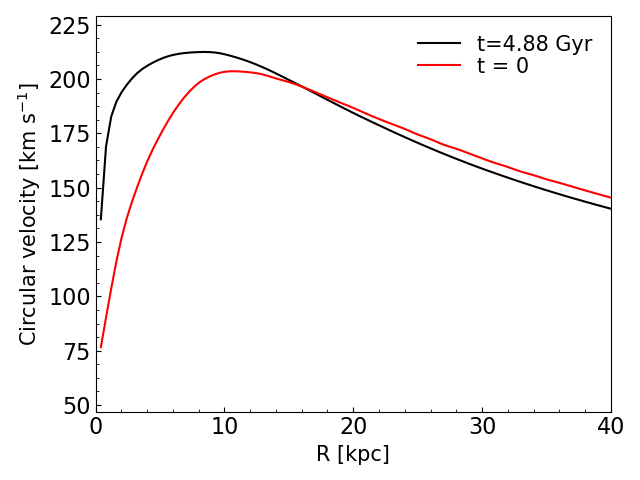}
	\includegraphics[width=0.645\columnwidth]{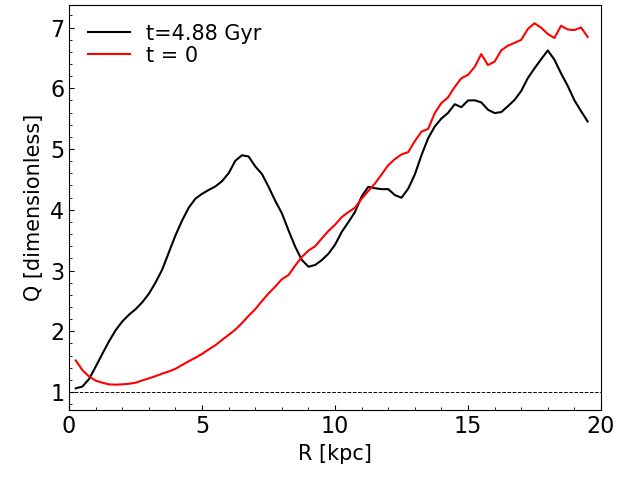}
	\includegraphics[width=0.645\columnwidth]{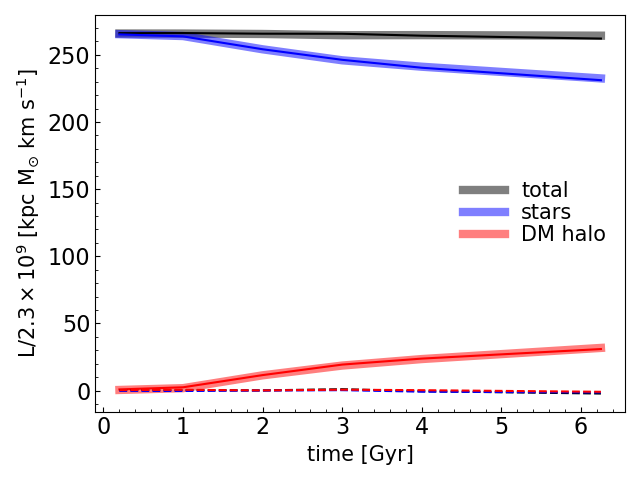}
	\caption{{\bf Left:} The initial rotation curve (red) and that at $t=4.88$ Gyrs, a snapshot where the double X/P configuration is well in place. {\bf Middle:} The Toomre $Q$ parameter calculated as a function of distance from the centre, at the same times as in the left panel. {\bf Right:} Angular momentum exchange between the stellar component and the dark matter halo, in the direction perpendicular to the disc (thin, solid), in the disc plane (thin, dashed) and total (thick).}
	\label{fig:vc_q_l}
\end{figure*}

The galaxy is evolved for a total of 7 Gyr in isolation, during which time it undergoes a series of secular dynamical processes which give rise to the key structural components of interest in this study. By $t \approx 5$ Gyr the stellar bar is well in place and a double, nested X/peanut configuration has formed and begins to stabilise into a steady structure symmetric about the disc plane. The initial rotation curve is shown in Figure \ref{fig:vc_q_l} (left panel) and peaks at roughly 10 kpc, beyond which it displays a steady shallow decline. Overplotted in the same panel is the rotation curve at $t=4.88$ Gyr, which shows a much steeper inner ascent, a broader peak and a slightly steeper decline compared to the initial conditions. The middle panel displays the disc's gravitational stability to fragmenting, as expressed through the Toomre $Q$ parameter (\citealt{Toomre1964}; i.e., stable for $Q >$ 1) calculated as a function of radius, initially, and at $t = 4.88$ Gyr as before. Finally, as the disc evolves it transfers steadily angular momentum to the live halo, as shown in the right panel of Figure \ref{fig:vc_q_l}. This is rather typical, and by $t=6.45$ Gyr the halo takes away roughly 13\% of the disc's angular momentum, essentially entirely from the vertical component $L_z$ (i.e., the rotation).

\begin{figure*}
	\centering
	\includegraphics[width=0.975\textwidth]{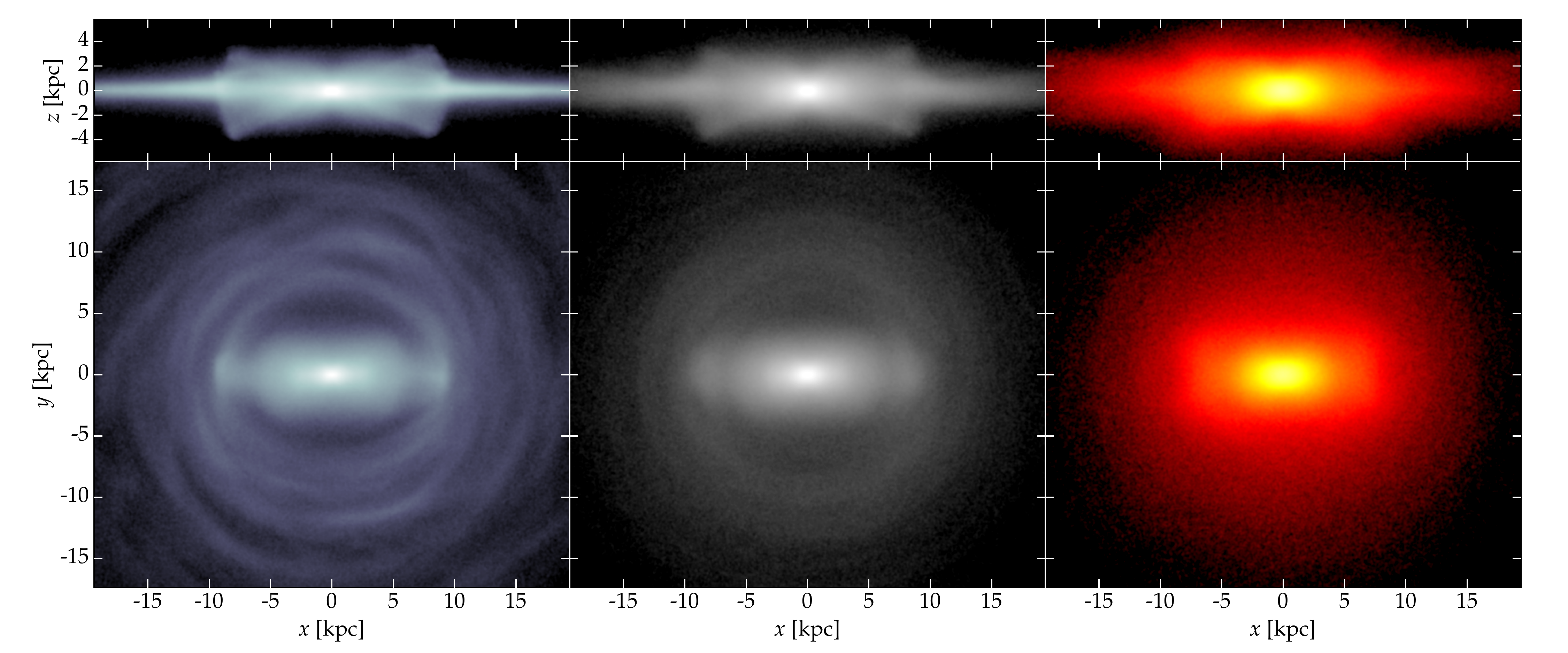}
	\caption{The three kinematic components after 6.45 Gyrs: cold/thin (left), warm/intermediate (middle) and hot/thick (right). Top panels represent and edge-on view with the bar aligned such that it is side-on, while bottom panels represent the face-on view. The ``outer'' peanut reaches its maximum vertical extent at roughly the location of the ansae, as seen in the cold and intermediate components.}
	\label{fig:sim145_proj_kin}
\end{figure*}

Figure \ref{fig:sim145_proj_kin} displays separately the 3 kinematical components of the simulation after $t= 6.45$ Gyr. In particular, the face-on view of the cold component -- which is the most sensitive to dynamical instabilities and perturbations (see e.g., \citealt{Fragkoudi+2017b}) reveals that the disc has formed multiple spiral arms outside $\sim$ 10 kpc, while within this radius there is a clear ``thick'' bar which has a lozenge or box-like shape, with a radius of roughly 6 kpc. 

Beyond this radius, a thinner and dimmer bar extends outwards to approximately 10 kpc, and ends in ansae, or bar ``handles'' as they are sometimes referred to. These features are symmetric density enhancements at the ends of the bar, quite common in old, evolved systems like barred lenticulars \citep{Martinez-Valpuesta+2007}. While the ansae are still discernible in the intermediate kinematic component, they are mostly ``washed out'' in the hot component, which retains the ``thick'' bar as the only distinctive structure apart from the disc. This progressive sensitivity to dynamical instabilities, from thin to thick components, will prove useful in disentangling the complicated structure of the bar in this simulation, and its associated X/P structures, as we detail in the following.

\section{Analysis}\label{sec:analysis}

In this Section we dissect the simulation snapshot 145 corresponding to 6.45 Gyrs of evolution from the initial conditions, an epoch chosen such that the double X/P configuration of interest is well in place and stable. Mock-images were generated by projecting the simulation particles in the ($x-y$) plane (disc viewed face-on) and ($R-z$) plane (disc viewed edge-on), with $R$ being the radial co-ordinate along the direction of the bar's major axis. This was done for all stellar particles as well as in detail for each kinematic component displayed Figure \ref{fig:sim145_proj_kin}. 

The edge-on view of this galaxy, as seen in the top panels of Figure \ref{fig:sim145_proj_kin}, reveals the presence of a significant X--shaped structure extending out to, and indeed reaching a maximum at, a radius just under $\sim$ 10 kpc. This is somewhat surprising, for two reasons. Firstly, this scale corresponds to roughly the end of the ``thin'' bar we see in the face-on view, whereas most of the studies to date find that it is only the inner half of bars which become unstable, and buckle to form a peanut (\citealt{Lutticke+2000}, \citealt{Skokos+2002a}, \citealt{Skokos+2002c}, \citealt{Laurikainen&Salo2017}, \citealt{Erwin&Debattista2017}). Secondly, this scale coincides with the location of the ansae, which are generally believed to be confined to the disc plane. Apart from this apparent X--shaped structure, one can just discern a similar (though shorter) feature nested within. This inner peanut extends out to a radius just inside $\sim$ 4 kpc, and also appears to be more boxy/peanut in shape, as opposed to the larger structure, which is rather strongly X--shaped. In the following we identify and quantify in detail the structural constituents of this simulated galaxy, separately for each disc component as well as their cummulative surface density distribution. 

\subsection{Analysis of the Surface Density Distribution}

\subsubsection{Structural Decomposition}

\begin{figure}
	\centering
	\includegraphics[width=0.975\columnwidth]{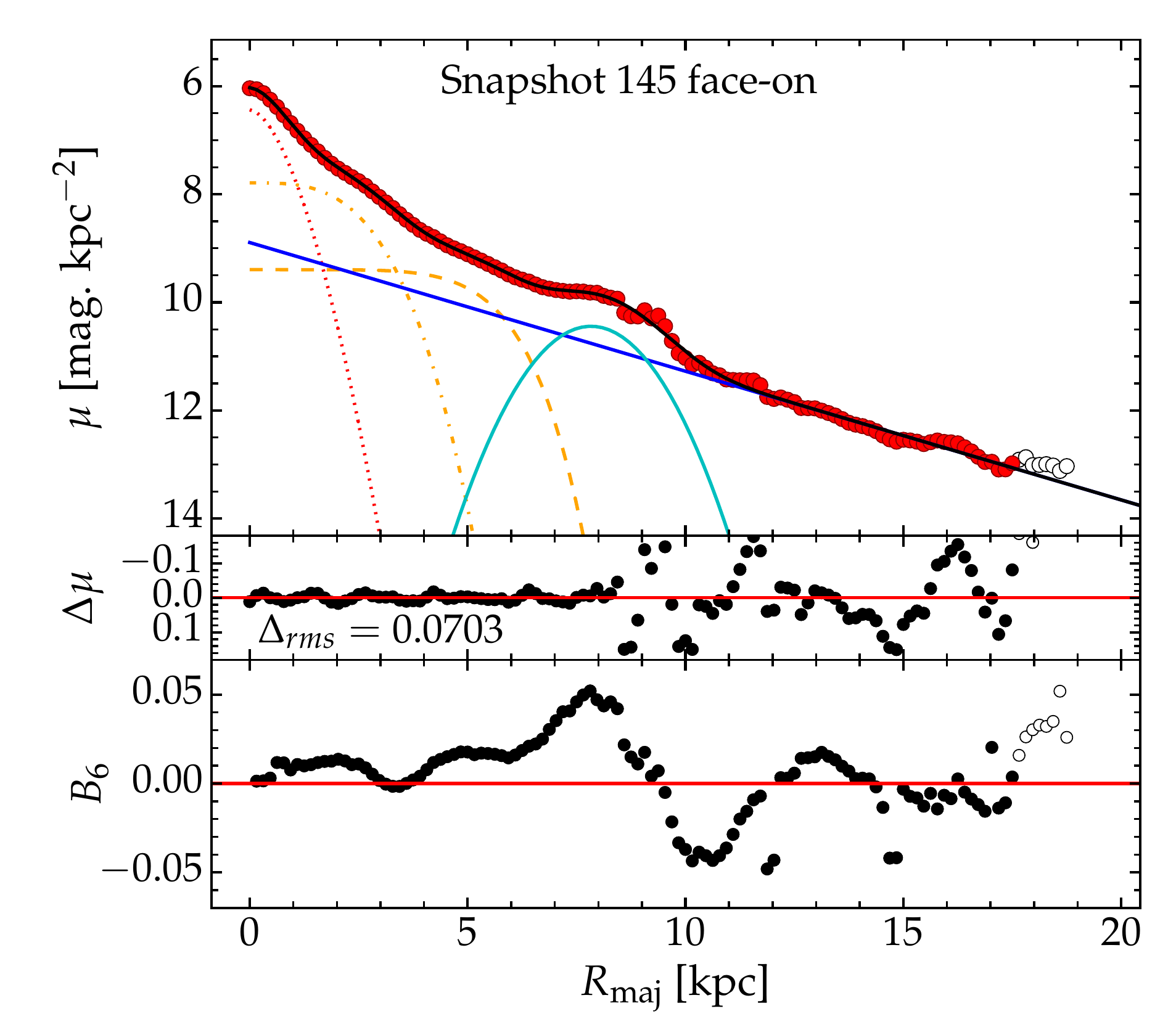}
	\caption{Structural decomposition of the surface brightness profile as observed in the simulation after 6.45 Gyr, in face-on orientation. {\bf Top:} The best-fit decomposition consisting of an exponential disc (blue), a Gaussian ring (cyan), and three S\`ersic components representing the pseudobulge (inner-most, red), and the two `nested' bars (orange). {\bf Middle:} Residual profile (data minus model). {\bf Bottom:} Radial $B_6$ profile showing a `dip' after $\sim$ 10 kpc, indicating the presence of ansae at that scale \citep{Saha+2018}.}
	\label{fig:decomp_145}
\end{figure}

We proceeded to analyse the mock-images described before as though they were real observations. The goal always being to perform a direct comparison between the simulation and galaxy observations, with the same methods and tools, we started by investigating the surface density distribution in these mock-images, mirroring the analysis of surface $brightness$ distributions of galaxies in astronomical images. In the latter case, the radial surface brightness profile is routinely extracted, by fitting the galaxy isophotes as a function of photocentric distance, and further decomposed into its structural constituents (disc, bulge, bar, etc.). Here we extracted the radial surface density profile from the simulation in the same manner. We employed the software {\sc Isofit}\footnote{\url{https://github.com/BogdanCiambur/ISOFIT}} \citep{Ciambur2015} to fit quasi-elliptical iso-density contours, as a function of distance (semi-major axis) from the centre, onto the mock-image oriented in face-on view. 

\begin{table}
%\begin{minipage}{30mm}
\caption{Decomposition Parameters}
\centerline{
	\begin{tabular}{l c c c c}
	  \hline 
	  Compt. [fn.$^{(a)}$] & $\mu_0/\mu_e$  & $h/R_e/R_{\rm ansa}$ & $W^{(b)}_{\rm ansa}$ &  $n$\\
	   & [log counts] &  [kpc] & [kpc] & \\
	  \hline \hline \\
	   Disc [E] & 8.94 & 4.88 & -- & --\\
	   Ansa [G] & 10.35 & 7.89 & 2.59 & --\\
	   Thin bar [S] & 9.76 & 4.26 & -- & 0.11\\
	   Thick bar [S] & 8.01 & 2.13 & -- & 0.40\\
	   Pseudobulge [S] & 7.41 & 0.79 & -- & 0.56\\
	  \hline 
	\end{tabular}\\	
}
\label{tab:decomposition}
$(a)$-- analytical function used to model the component:\\
 E = exponential; G = Gaussian; S = S\'ersic \\
$(b)$-- full width at half-maximum of the Gaussian component\\
%\end{minipage}
\end{table}

A structural decomposition of the resulting one-dimensional surface density profile was performed with the software {\sc Profiler}\footnote{\url{https://github.com/BogdanCiambur/PROFILER/}} \citep{Ciambur2016}, and the resulting best-fit model is shown in the top panel of Figure \ref{fig:decomp_145}. The outer profile is dominated by the disc, which was modelled with an exponential. At a scale of $9$ kpc, an excess from this exponential becomes apparent, and corresponds to the ansae. This was modelled with a Gaussian, which is the common parametric function employed to fit rings. Interior to this, the thin and thick bars were each modelled with S\'ersic functions with indices ($n$) less than 1. Finally, a final excess becomes apparent in the central kpc, which is the pseudobulge that has developed from this galaxy's secular evolution. This was also modelled with a S\'ersic function. The parameters of the best-fit solution are shown in Table \ref{tab:decomposition}. The measured disc scale length at this epoch shows insignificant (less than 5$\%$) evolution from the initial conditions, specifically compared to the thin disc $h_{\rm ini}=4.7$ kpc, which dominates the outer profile. As a cautionary note, these best-fit  parameters of course correspond to the azimuthally-averaged 1D profile along a single dirrection, that of the major axis of the iso-density contours, i.e., approximately along the major axis of the bar (although the ellipse position angle was allowed to vary with radius). Although a simulation has the advantage of providing a complete 3D view of the galaxy, we again stress that the intention is to perform a one-to-one comparison of its structural properties with those in real galaxies, i.e., derived with the same method (that employed in CG16) in both cases.

The bottom panel in Figure \ref{fig:decomp_145} shows the radial $B_6$ profile as extracted in the face-on view. While the edge-on orientation is of interest for computing the X/P parameters, it is worth noting that even in face-on view, the $B_6$ signal is positive and significant along the length of the bar components, which implies that the morphology remains boxy/peanut/X--shaped at all inclinations. Although multiple studies have shown that, in general, buckled bars display X/P isophotes at high disc inclination, they gradually assume the morphology of a ``barlens'' (\citealt{Laurikainen+2011_barlens}, \citealt{Athanassoula+2015}, \citealt{Salo&Laurikainen2017}) as they approach face-on view. Nevertheless, \cite{Saha+2018} have shown through collisionless simulations that a 3--dimensional peanut morphology can indeed develop, supporting the earlier orbital analysis work of \cite{PatsisKatsanikas2014}, who show that orbits supporting sharp X--like features located at energies beyond the ILR (non-periodic) might have a double-boxy nature, i.e., both in face-on and side-on views. Indeed, recently \cite{Smirnov&Sotnikova2018} have shown that the presence of a bulge component, higher  parent disc thickness, or higher Toomre $Q$ parameter, are all factors which determine a barlens morphology in-plane. Consistent with these studies, our bulge-less simulation displays a 3D peanut, albeit with a weaker X/P signature in face-on view -- where isophotes appear rather boxy, than that in edge-on projection -- where the isophotes appear strongly X/peanut--shaped. The face-on $B_6$ profile suddenly turns negative at, and just after, the scale of the ansae, a  behaviour additionally noted in \cite{Saha+2018} and interpreted as a tell-tale sign of the presence ansae. 

The surface density profile thus indicates the presence of an inner, thick bar, as well as a longer and thinner bar component. This is not uncommon, and indeed shows similarities with the light profiles of NGC 2549 and, to a lesser extent, NGC 128, the two galaxies found to show a double peanut signature in CG16. We refer the reader to Figs. 4 and 9 in their paper, for comparison. Furthermore, the short, thick bar, and longer, thinner bar evident in this simulation may constitute a configuration similar to the Milky Way (\citealt{Martinez-Valpuesta&Gerhard2011}). The presence of a long thin bar, in addition to the classical bar of our Galaxy, was noted by \cite{Hammersley+1994}, based on star counts in the Galactic plane. Further studies employing more refined tracers (Red Clump giants) have since confirmed its presence, and led to a series of papers where it is debated whether this long component is coupled to, or is misaligned with, the ``classical bar'' or ``triaxial bulge'' as it is sometimes referred to (\citealt{Hammersley+2000}, \citealt{Lopez-Corredoira+1999,Lopez-Corredoira+2007}, \citealt{Cabrera-Lavers+2008}, \citealt{WeggGerhard&Portail2015}). 

\subsubsection{Quantifying the X/Peanut Structures}\label{subsec:peanut_params}

Apart from the surface brightness profile along the major axis of the quasi-ellipses, {\sc Isofit} also provides the geometric parameters of the iso-density contours, namely their position angle, ellipticity, and deviations from pure ellipses. These deviations are important, as they contain physical information pertaining to the different structures that make up the galaxy. Briefly, they are calculated by first fitting a pure ellipse on the mock image, at a given radius from the centre, and then performing a Fourier expansion along the ellipse, such that

\begin{equation}
\rho(\psi) = \rho_0 + \sum_{n} [A_n{\rm sin}(n\psi) + B_n{\rm cos}(n\psi)],
\end{equation}

\noindent where $\rho_0$ is the best-fit``average'' surface density of the pure ellipse, $\psi$ is the eccentric anomaly angle which defines a point on this ellipse, $\rho(\psi)$ is the surface density at that point, while $n$ and $A_n$, $B_n$ are the order and coefficients of the different Fourier harmonics in the expansion. We refer the reader to \citep{Ciambur2015} for further details. In the past, the $B_4$ harmonic has been used extensively to quantify the ``boxyness'' or ``discyness'' of galaxy isophotes, and correlating it with the galaxy's physical properties (e.g., \citealt{Davies+1983}; \citealt{Bender1988}; \citealt{Nieto+1991}; etc.). More recently, CG16 have shown that the $B_6$ harmonic can serve as a robust tracer of X/peanut structures, and can serve to obtain quantitative parameters such as the length, height above the disc plane, and integrated ``strength'' of the instability. The $B_6$ was used for this purpose to study the geometry of the Milky Way's own X/peanut-shaped structure and bar \citep{CiamburGraham&Bland-Hawthorn2017}. Further, \cite{Saha+2018} have applied this method to study the face-on view of X/P structures in simulations. In this work, the radial $B_6$ profile will serve as the primary tracer of the X/P feature. 

\begin{figure}
	\centering
	\includegraphics[width=0.975\columnwidth]{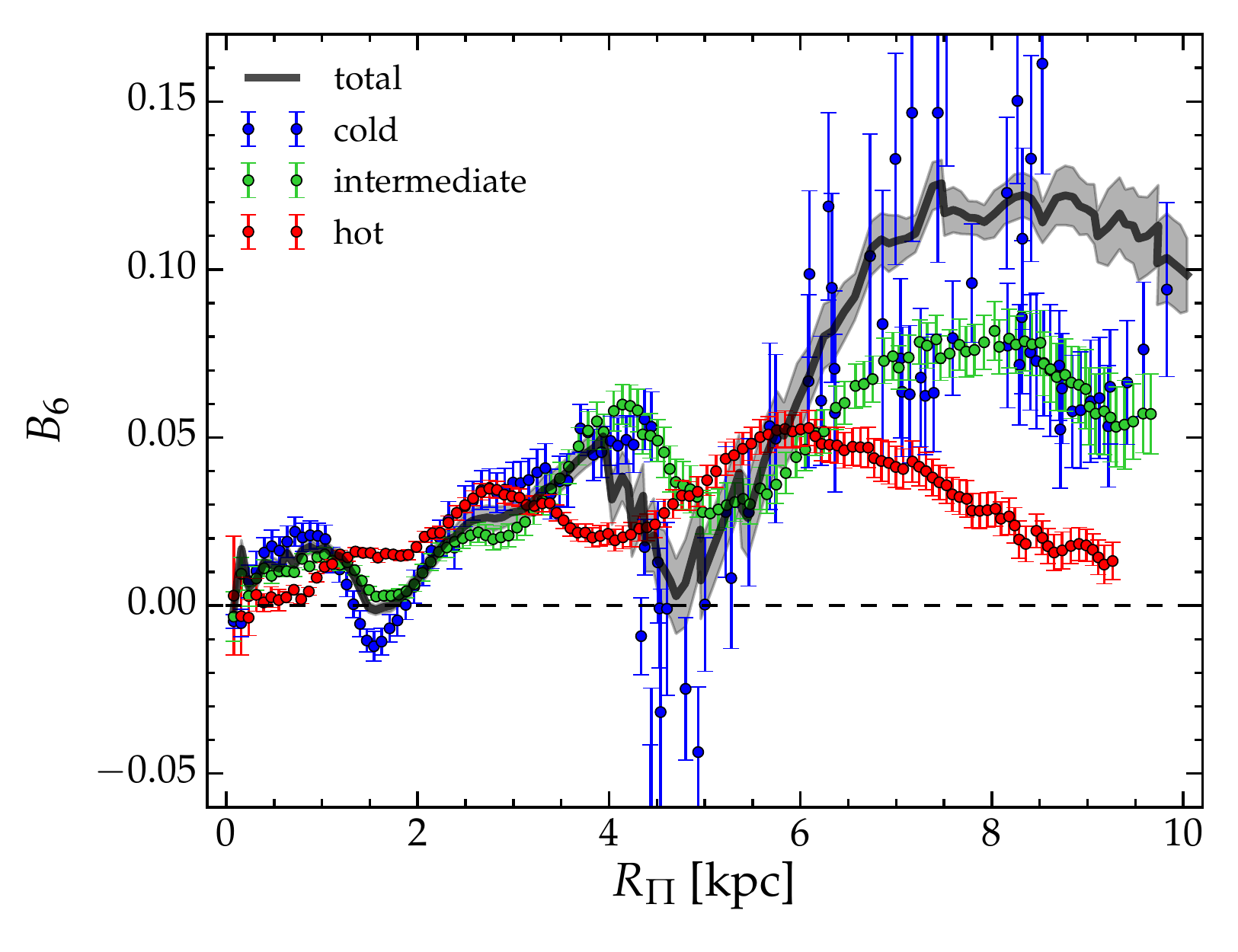}
	\caption{The X/P signature as measured in the 3 different kinematic components, as well as their superposition. (Cold component dominates the total signal, hot component is offset radially).}
	\label{fig:B6_kin}
\end{figure}

\begin{table*}
\begin{minipage}{60mm}
\caption{The X/P Diagnostics}
\centerline{
	\begin{tabular}{l c c c c c c}
	  \hline 
	  & ${\it \Pi}_{\rm max}^{(a)}$ & $R_{{\it \Pi},{\rm int}}^{(b)}$ & $z_{{\it \Pi},{\rm int}}^{(c)}$ &  $S_{\it \Pi}^{(d)}$ & $W_{\it \Pi}^{(e)}$ & $\psi^{(f)}$  \\
	   & & [kpc,  units of $h$] &  [kpc, units of $h$] & [kpc, units of $h$] & [kpc, units of $h$] &[$\degree$]  \\
	  \hline \hline \\
	  In/thin & 0.056$\pm$0.010 & 4.27$\pm$0.82, --- & 1.02$\pm$0.82, --- & 8.06$\pm$2.91, --- & 1.95$\pm$0.29, --- & 13 \\
	  In/interm. & 0.060$\pm$0.006 & 4.14$\pm$0.82, --- & 1.63$\pm$0.82, --- & 7.57$\pm$3.09, --- & 1.62$\pm$0.29, --- & 22 \\
	  In/thick & 0.035$\pm$0.003 & 2.75$\pm$0.82, --- & 1.64$\pm$0.82, --- & 5.59$\pm$2.04, --- & 2.08$\pm$0.29, --- & 31 \\
	  {\bf In/total} & 0.050$\pm$0.006 & 3.94$\pm$0.21, 0.89$\pm$0.11 & 1.45$\pm$0.21, 0.33$\pm$0.11 & 5.15$\pm$2.21, 1.17$\pm$0.50 & 1.49$\pm$0.29, 0.34$\pm$0.09 & 20\\
	  \hline
	  Out/thin & too noisy & --- & --- & --- & --- & --- \\
	  Out/interm. & 0.082$\pm$0.009 & 8.03$\pm$0.82, --- & 2.45$\pm$0.82, --- & 25.51$\pm$6.57, --- & 3.81$\pm$0.58, --- & 17 \\
	  Out/thick & 0.053$\pm$0.005 & 6.09$\pm$0.82, --- & 2.56$\pm$0.82, --- & 14.45$\pm$4.05, --- & 3.46$\pm$0.58, --- & 23 \\
	  {\bf Out/total} & 0.122$\pm$0.012 & 8.33$\pm$0.22, 1.88$\pm$0.21 & 2.44$\pm$0.21, 0.55$\pm$0.20 &  --- & --- & 16 \\
	  \hline 
	\end{tabular}\\	
}
\label{tab:diagnostics}
\end{minipage}
\begin{minipage}{169mm}
$(a)$-- maximum amplitude of $B_6$ harmonic; $(b)$-- intrinsic radius of X/P structure; $(c)$-- intrinsic vertical height of X/P structure; $(d)$-- integrated strength of the $B_6$ profile; $(e)$-- full width at half-maximum of the $B_6$ profile; $(f)$-- angle of the ``X'' arm relative to major axis.
\end{minipage}
\end{table*}

To keep equivalence with the observational sample of CG16, we extracted the radial $B_6$ profile from the mock-images corresponding to projections of the simulation particles where the disc is viewed edge-on. In addition, the bar was also rotated such that it is viewed side-on. The $B_6$ profile was extracted for each disc component individually, as well as cumulatively from the superposition of the 3 components. The result is shown in Figure \ref{fig:B6_kin}. It is immediately apparent in this figure that the total signal, plotted as the thick black curve, shows a prominent peak at roughly 8 kpc, which corresponds to the large-scale X--shaped structure clearly discernible in Figure \ref{fig:sim145_proj_kin}. This scale additionally coincides with the ansae signature in the density profile (the Gaussian component in Figure \ref{fig:decomp_145}, which peaks at $R_{\rm ansa}=7.89$ kpc). The second feature of note is a second, intermediate--strength peak at an interior radius, of $\sim$ 4 kpc. This corresponds to the smaller, more peanut-shaped feature nested within the large-scale X. When we dissect the total signal by kinematic components, we find, unsurprisingly, that the thin component dominates the outer X/P signature, and also shows a strong signal for the inner peanut. The intermediate component is comparable in amplitude at 4 kpc, but sub-dominant (though still significant), for the outer X. Finally, the hot component also appears to show two significant peaks, less prominent than in the colder components, and offset towards the centre with respect to the intermediate and the cold components' $B_6$ signatures. Interestingly, we note a third peak in the $B_6$ profile occurring at $R_{\Pi} \approx 0.6$ kpc in the cold and warm components, and just beyond 1 kpc in the hot disc, where it is not clearly separated from the other peaks at larger radii. This could indicate the presence of a third peanut structure, possibly associated with a decoupled nuclear bar. Indeed, the beat feature which appears in the main bar's strength ($A_2$) and pattern speed evolution, as shown in Figure \ref{fig:omega_p_vc}, could result from its interferecne with such a nuclear bar. This third inner  X/P feature tentatively shows up in the face-on density projections as well, but occurs on a scale too small for a fully reliable measurement. Hence, we do not include this feature in our analysis. 

As per CG16, we compute the different parameters of the X/P structures, including the intrinsic radius ($R_{\Pi, {\rm int}}$), the height above the disc plane ($z_{\Pi, {\rm int}}$), the integrated ``strength'' ($S_{\Pi}$), starting from the $B_6$ profile. Briefly, these correspond to the radial scales where $B_6$ reaches a maximum\footnote{NB: This location does not necessarily correspond to the ``end'' of the peanut, as the $B_6$ usually falls-off more or less gradually from the maximum point. Hence, a weaker X/P signature remains, and trails off, beyond the $B_6$ peak location.} for the former two, while the latter is defined as the integrated $B_6$ profile within the FWHM around the peak. The full set of parameters is presented in Table \ref{tab:diagnostics}.

\subsection{A Few Notes on the Dynamics}

\begin{figure}
	\centering
	\includegraphics[width=0.975\columnwidth]{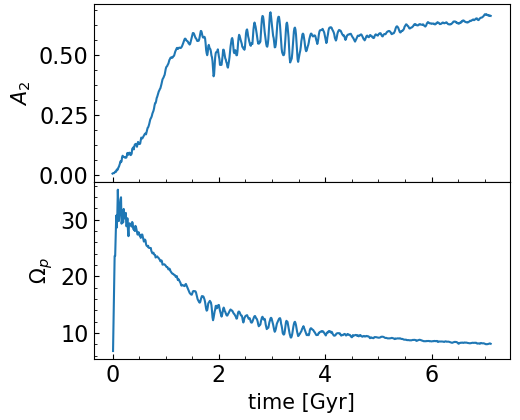}
	\caption{The bar strength ({\it top}) and pattern speed ({\it bottom}) as a function of time.}
	\label{fig:omega_p_vc}
\end{figure}

Before attempting to interpret the double X/P configuration, there are a number of aspects that are worth pointing out. First, the fact that the outer X overlaps the face-on ansae appears surprising, since the latter are canonically believed to accumulate around the co-rotation radius, where vertical instabilities are minimal. However, this is a relatively slow bar, rotating with a pattern speed of $\Omega_p \approx 9$ km/s/kpc (see Figure \ref{fig:omega_p_vc}), which places the co-rotation radius beyond 20 kpc, almost twice the scale of these ansae. \cite{Skokos+2002b} have studied the parameter space of buckling bars, including the bar pattern speed, and have shown that slow bars are more prone to develop an inner secondary structure which might be peanut-like, though in their case is oriented along the minor axis of the main bar.

Second is the fact that the bar extends to only $\sim$ 1/2 the radius of co-rotation. While surprising, this might be the effect of a tangential shear induced by the rather sharply-declining rotation curve beyond 10 kpc (roughly the end of the bar), as shown in Figure \ref{fig:vc_q_l}, which may prevent the bar to extend beyond that radius and possibly entrain the formation of the ansae as well as confine the outer X--shaped structure. Indeed, the galaxy rotation curve has been shown to affect the 3D morphology of X/P structures (\citealt{Salo&Laurikainen2017}). A similar example, of stellar overdensities accumulating at the end of the bar are the so-called ``spurs'' observed in X/P galaxies in nearly, though not completely, edge-on orientations, often beginning to wind around for a short spatial extent as they emerge from the bar's two ends. By contrast ansae are symmetric, not showing any particular winding direction. While X/P galaxies with spurs show no clear vertical signature at the scale of the spurs, we note that this is the case for this simulation as well. In particular, the vertical extent of the outer X only becomes apparent in edge-on view, and has no obvious manifestation at other orientations, as can be seen in Figures \ref{fig:sim145_proj_kin} and \ref{fig:anim1}. A study of the rotation curves of X/P galaxies with spurs would provide more insight towards this scenario.

\begin{figure}
	\centering
	\includegraphics[width=0.975\columnwidth]{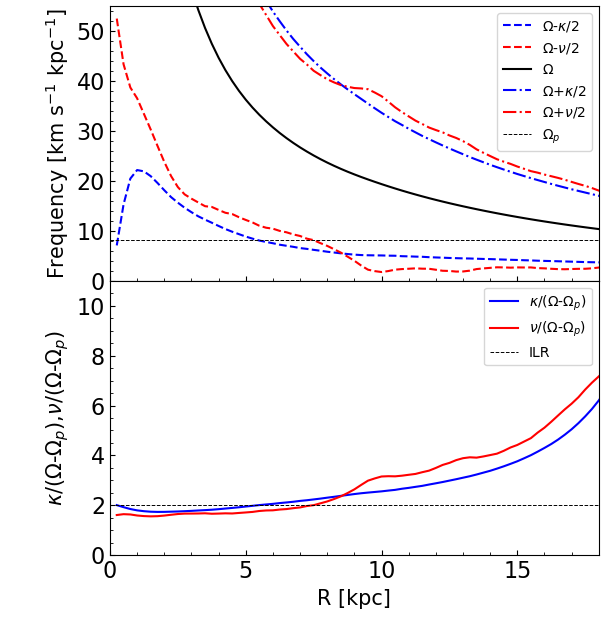}
	\caption{{\bf Top:} Angular velocity diagram of the simulation, after 6.45 Gyrs of evolution. $\Omega$ is the average stellar rotation, $\Omega_p$ is the pattern speed, and $\kappa$ and $\nu$ are the radial and vertical oscillation frequencies, respectively.  ($\Omega-\kappa/2$) and ($\Omega-\nu/2$) cross $\Omega_p$ at the radial and vertical Inner Lindblad Resonances, respectively. Co-rotation occurs at $R \sim 20$ kpc. {\bf Bottom:} Resonance diagram constructed from axisymmetric frequency ratios in the plane (blue) and vertical (red) directions.}
	\label{fig:omega}
\end{figure} 

In the top panel of Figure \ref{fig:omega} we display the angular velocity diagram calculated from the simulation snapshot of interest, i.e., $t=6.45$ Gyr, while the bottom panel shows a resonance diagram analogous to Fig. 8 in \cite{Combes+1990}. Here $\Omega$ is the circular orbit frequency, $\Omega_p$ the pattern speed of the bar, and $\kappa$ and $\nu$ the radial and vertical epicyclic frequencies, respectively. The crossing points between the $\Omega - \kappa/2$ and $\Omega - \nu/2$ curves with $\Omega_p$ define the radial and horizontal 2:1 inner Lindblad resonances, i.e., they correspond to two in-plane, and vertical, oscillations per one bar rotation, in the bar frame. In our specific case the two locations define a rather broad resonance region spanning in from $\approx 5.5 - 7.5$ kpc, a region which has been shown to be associated with periodic and quasi-periodic orbits supporting peanut-like features, and thus define (the vertical ILR in particular) the radial scale where such features extend the most out of the disc plane (\citealt{Combes+1990}, although see e.g., \citealt{PatsisKatsanikas2014}, \citealt{Patsis&Harsoula2018}, \citealt{Parul+2020} for discussions on non-periodic orbits around the ILR, supporting sharp X--shaped structures). Indeed, the vertical ILR location is broadly consistent, though slightly shorter than, the radial extent of the larger X structure, $R_{\rm \Pi} = 8.33\pm0.22$ kpc as measured from the cumulative mock-image (all 3 disc components). We return to this point in the following Section.

\section{Discussion}\label{sec:discussion}

In this section we discuss the development of the double X/P configuration in our simulated galaxy with an emphasis on the dynamics and orbital properties which are likely to support it, and further place it in context by comparing with the observational sample of CG16, consisting of twelve nearby galaxies with known peanut ``bulges'', of which two were found to host double, or nested, X/P structures.

\subsection{How Does This Double Configuration Arise?}

\begin{figure}
	\centering
	\includegraphics[width=0.9\columnwidth]{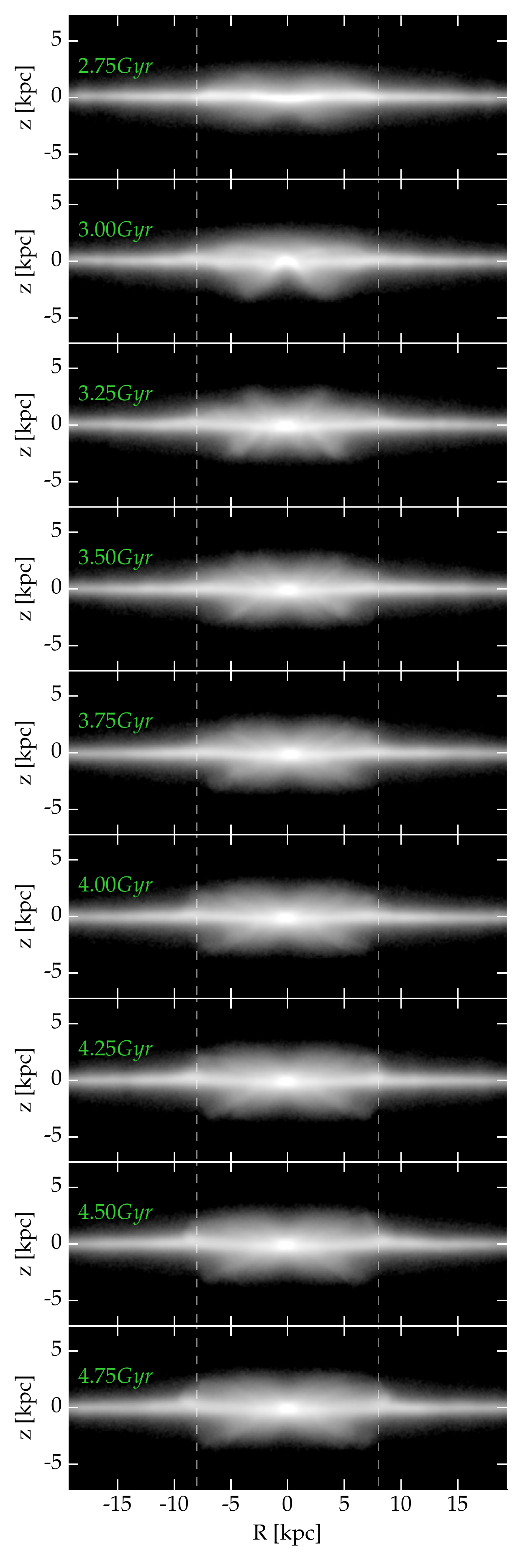}
	\caption{Progressive formation of the double X structures in the simulation. The dashed line marks 8 kpc, roughly the scale of the outer X structure once it has stabilised.}
	\label{fig:evolution}
\end{figure}

Back-tracing the simulation to $t \approx$ 2.75 Gyr, when the bar just starts to become unstable, we are able to observe in detail how the double-X structure arises and develops until the snapshot of interest. In Figure \ref{fig:evolution} we display multiple snapshots separated in time by 250 Myr, from the point where the bar begins to become vertically unstable, and up to the point where it begins to settle into the double X/P configuration ($t \approx$ 4.75 Gyr). 

Here we observe the buckling instability develop, characterised by vertically asymmetric bending modes (``smile -- frown", or ``banana" orbits), and rather strong within 3 -- 3.5 Gyr. This phase occurs mostly interior to 5 kpc, or roughly half the radius of the bar. By 3.5 Gyr we begin to notice 2 components in this thick part of the bar, namely a peanut-shaped component and an X-shaped feature over-lapping it, and slightly more extended in radius. During the subsequent evolution, the peanut-shaped structure undergoes very little change, and ends up as the inner peanut, whereas the X extends in radius and undergoes further buckling perpendicular to the disc plane. This component ends up as the strong outer X structure, peaking in strength at $\sim$ 8 and $\sim$ 7 kpc for the thin and intermediate components, respectively, and $\sim$ 6 kpc for the thick component (see Table \ref{tab:diagnostics}). By the epoch chosen for the analysis above (6.25 Gyr) the double configuration is well stabilised.

Throughout their evolutionary course, the two components retain their dynamical coupling, both rotating with the same pattern speed. They are thus clearly different manifestations occurring in the same stellar component, and therefore likely supported by different orbit families. Qualitatively, the setup appears to be a superposition of the two morphological classes of boxy/peanut/X --shaped structures defined by \cite{Bureau+2006}, namely ``CX'' and ``OX'', the first being features whose four X ``arms'' cross the photocentre ($\times$), while the second exhibiting arms which cross the disc plane but do not meet in the centre ($>$--$<$). By studying the orbital dynamics around the vertical ILR region, \cite{Patsis&Harsoula2018} have shown how a broad variety of X/Peanut morphologies can be supported through various combinations of stable 3D periodic orbits, and quasi-periodic or chaotic orbits trapped around them which can support either a CX or OX morphology. In particular, the general morphology of the outer ``X'' in our case -- bow-tie shaped, with a well defined and abrupt end -- is reminiscent of the OX configuration obtained in \cite{Patsis&Harsoula2018} via sticky-chaotic, or quasi-periodic orbits at different energies (their Figs. 5c and 7). The latter findings are indeed supported by \cite{Smirnov&Sotnikova2018} and \cite{Parul+2020}, who note the presence of double--X features in their simulation for certain initial conditions (in terms of Toomre $Q$ and disc scale height), where a secondary bar buckling eventually setlles into a stable double X--like configuration. It is possible that factors such as the absence of a significant classical bulge component to stabilise the inner bar, and the latter's relatively low pattern speed (cf. \citealt{Skokos+2002b}), play a role in facilitating the population of the different orbit families which end up supporting the two nested structures.

%\begin{itemize}
%\item Identify the snapshot whe each of the X/P structures form in the simulation -- explain formation of this configuration;
%\item Outer X/P still a buckling instability, as it has a smile-frown asymmetry episode in past;
%\item However, it coincides with ansae, extending al the way to the end of the bar;
%\item Still same structure, as all peanuts rotate with the same (bar's) pattern speed;
%\item Place all the X/P parameters on scaling relations plots observed for real galaxies \& Milky Way -- discuss the 2 observed double peanuts in context;
%\item Radial scale of the outer X/P structure corresponds to the ansae; Moreover, hot component does not show clear ansae, nor a well-defined outer X (although it does show a double peak in $B_6$, translated inwards). This points to an association between outer X and the ansae, which is OK since the angular velocity diagram in Figure \ref{fig:omega} indicates that corotation occurs well beyond, at $>$ 20 kpc;
%\item Figure \ref{fig:omega} also indicates that the inner peanut is a ``classical'' peanut structure which arises around the inner Lindblad resonance;
%\end{itemize}

\begin{figure}
\centering
    \animategraphics[width=.95\columnwidth,loop,autoplay]{9}{./frame-}{0}{20}
    \caption{Face-on to edge-on animation with the bar oriented at 45$\degree$ with respect to the line of sight to the centre. Evident in edge-on view, the strong X--shaped structure becomes indistinguishable at, and close to, face-on orientation.}
    \label{fig:anim1}
\end{figure}

%\item Main bow-tie peanut corresponds to ansae; Support using ILR location (from Paola)
%\item Comment on face-on decomposition and B6 dip corresp to ansae
%\item Make a point that peanut extends to end of the bar, not halfway - would have %higher-than -average RPI/h ratio: interpret?
%\item Initial conditions, combined disks have $h=4.42 \pm 0.52$ kpc
%\item Possibly truncated disc showing spiral arms due to bar orbit restructuring

\subsection{Comparison with Nearby Galaxies}

\begin{figure*}
	\centering
	\includegraphics[width=0.975\columnwidth]{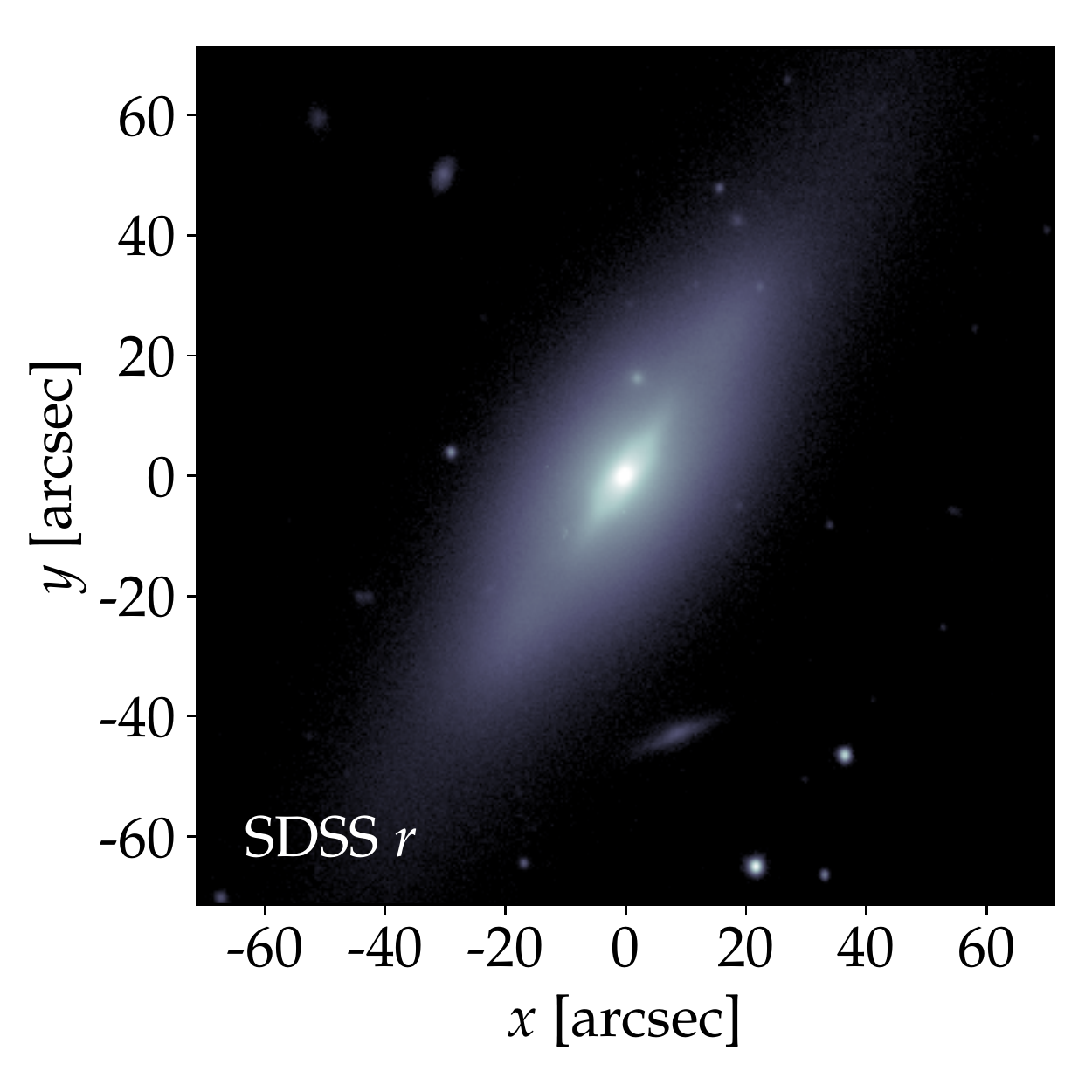}
	\includegraphics[width=0.975\columnwidth]{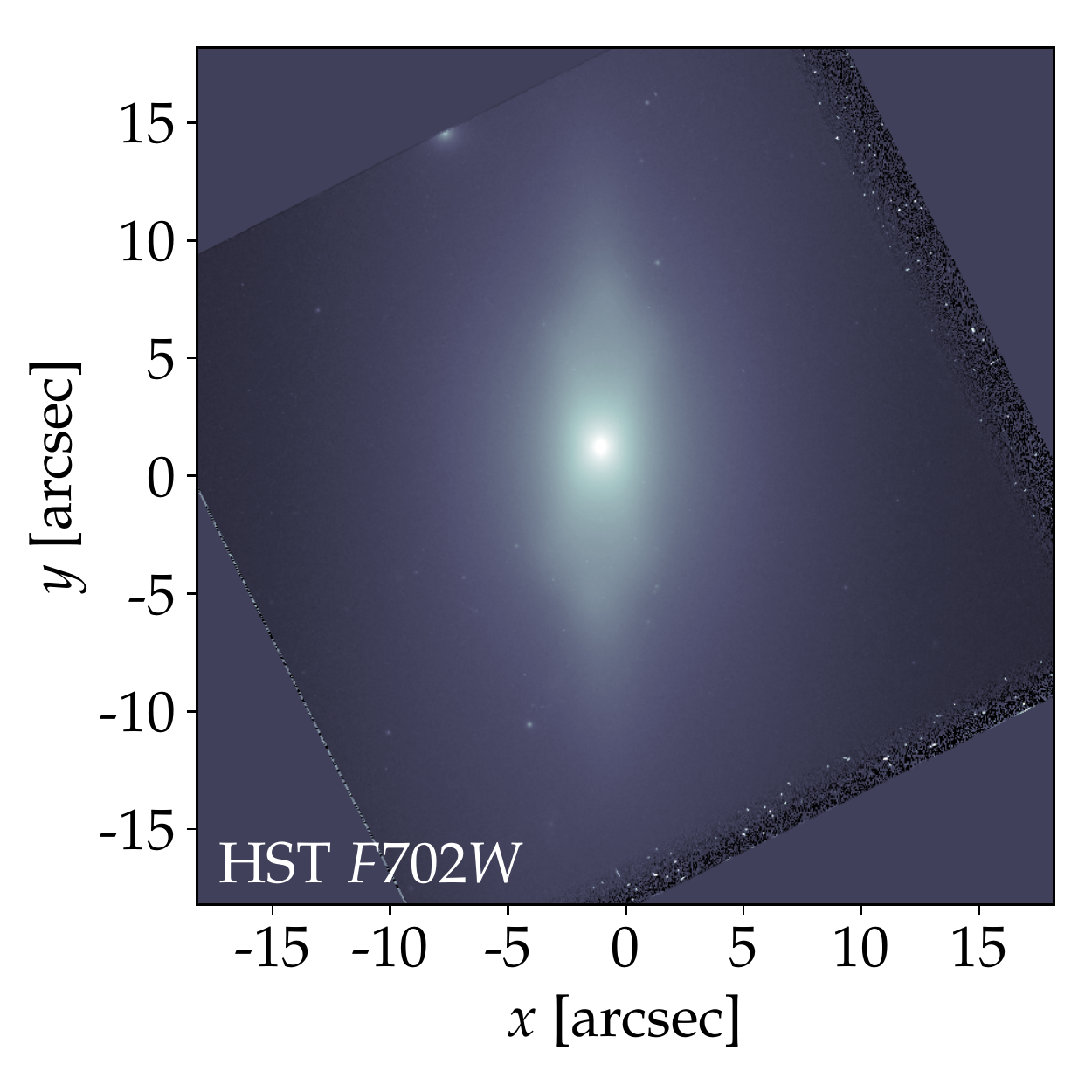}
	\caption{NGC~2549, an edge-on galaxy with a double X/P configuration, as observed with SDSS in the $r$ band (left), and at higher resolution with HST, at 702 nm (right). Unlike the simulation analysed in this paper, here the inner peanut structure appears to be stronger, with a well defined bow-tie, or `X' shape, while the larger scale structure is weaker. The two X/P structures are well separated, as shown by their extracted $B_6$ profile (see \citealt{Ciambur&Graham2016}, Figure A1 in their Appendix.)}
	\label{fig:ngc2549}
\end{figure*}

CG16 have studied the structural parameters of X/P structures in 12 edge-on galaxies known in the literature to host such features. They have shown that some of the peanut parameters -- as measured analogously here, in Section \ref{subsec:peanut_params} -- are correlated, roughly defining a number of scaling relations. These provide a useful benchmark to compare our simulation results against real galaxies, and indeed the Milky Way (\citealt{CiamburGraham&Bland-Hawthorn2017}). Of particular interest is the fact that two objects in the CG16 sample, NGC~128 and NCG~2549, were found to exhibit a double peanut signature, with a (at the time) unusually small\footnote{Unusually small for a large-scale bar plus embedded nuclear bar configuration.} scale ratio of $\sim$ 5:1 and $\sim$ 3:1 respectively. For comparison with our work, we show a large-scale observation of NGC~2549 in the SDSS $r$--band, and a higher-resolution zoom-in observation taken with the Hubble Space Telescope, at 702 nm, in Figure \ref{fig:ngc2549}. NGC~2549 can indeed be regarded as an observational analogue of our simulation, the two displaying many structural similarities (compare Figure \ref{fig:decomp_145} with the photometric decomposition of NGC~2549 shown in CG16, their Fig. 9). Although NGC~128 was also found to host nested peanut features in the aforementioned paper, we refrain to compare our simulation with this object, as this is a strongly perturbed, peculiar lenticular galaxy (S0 pec) with an offset, counter-rotating gas disc and a visibly warped stellar disc, discernibly interacting -- exchanging mass --  with its companion NGC~127 (\citealt{Emsellem&Arsenault1997} and references therein). The latter process very likely contributes to the peanut instability, and indeed leads to outlier peanut parameters, particularly strength.  

\begin{figure}
	\centering
	\includegraphics[width=0.975\columnwidth]{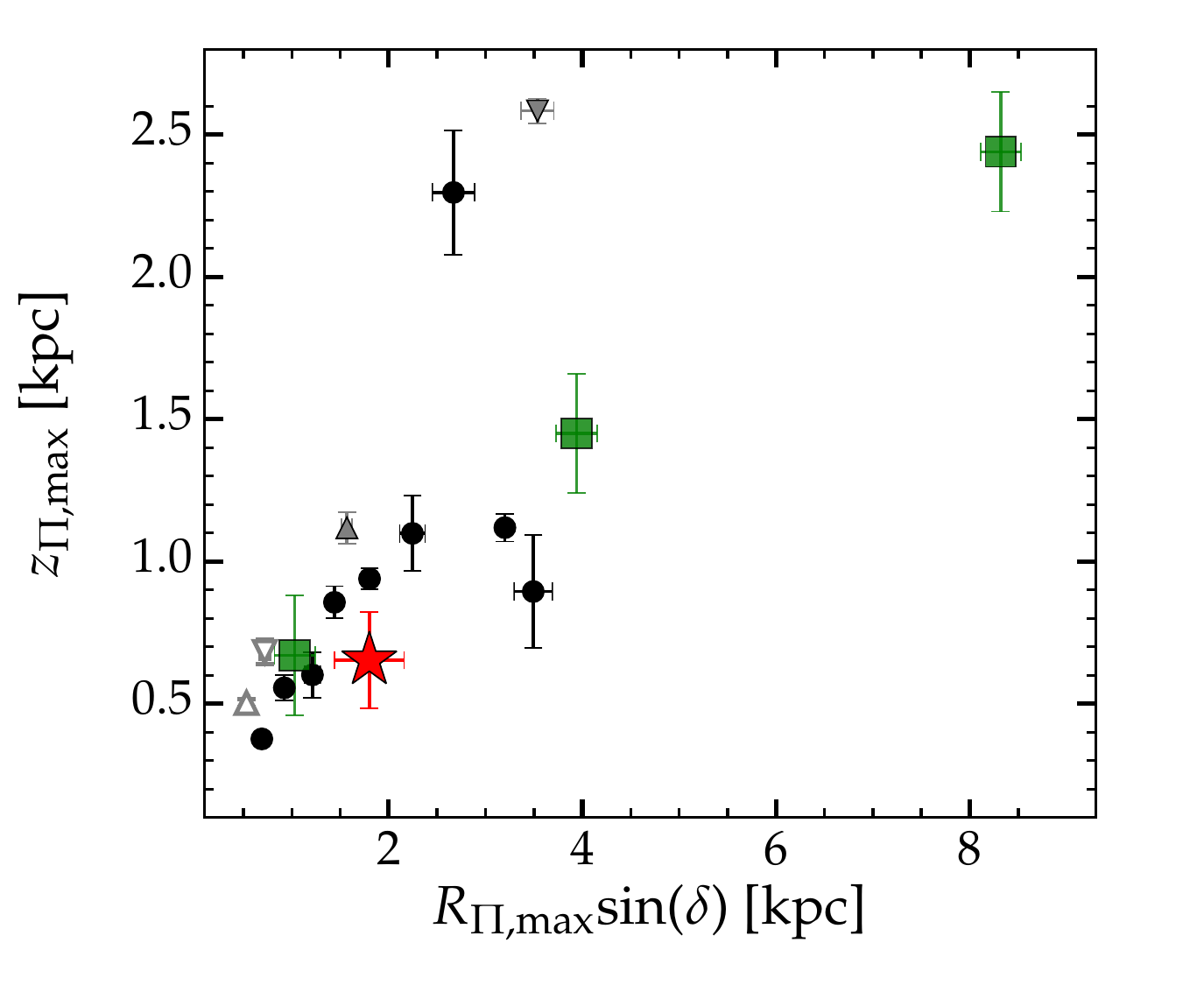}
	\caption{The vertical size ($z_{{\rm\Pi,max}}$) of the X/P structure as a function of its projected radial extent, $R_{\rm \Pi,max}{\rm sin}(\delta)$. Solid circles correspond to the observational sample in \citealt{Ciambur&Graham2016}, the star symbol corresponds to the Milky Way (with intrinsic, not projected, $R_{\rm \Pi,max}$), while the green squares correspond to the 3 peaks in the $B_6$ profile of this simulation.}
	\label{fig:R_vs_z}
\end{figure}

Figure \ref{fig:R_vs_z} displays the correlation between the observed radii of X/P structures and their vertical extent above the plane of the disc. The grey symbols correspond to NGC~128 (downward triangles) and NGC~2549 (upward triangles), filled for the outer and empty for the inner peanut, respectively. The red star corresponds to the Milky Way, while the green squares correspond to the X/P features (inner and outer) studied in this work. Although the outer feature is quite clearly larger than the sample average, notably in radius, both the inner and outer X/P structures lie on the general trend nevertheless. 

\begin{figure*}
\includegraphics[width=.9\columnwidth]{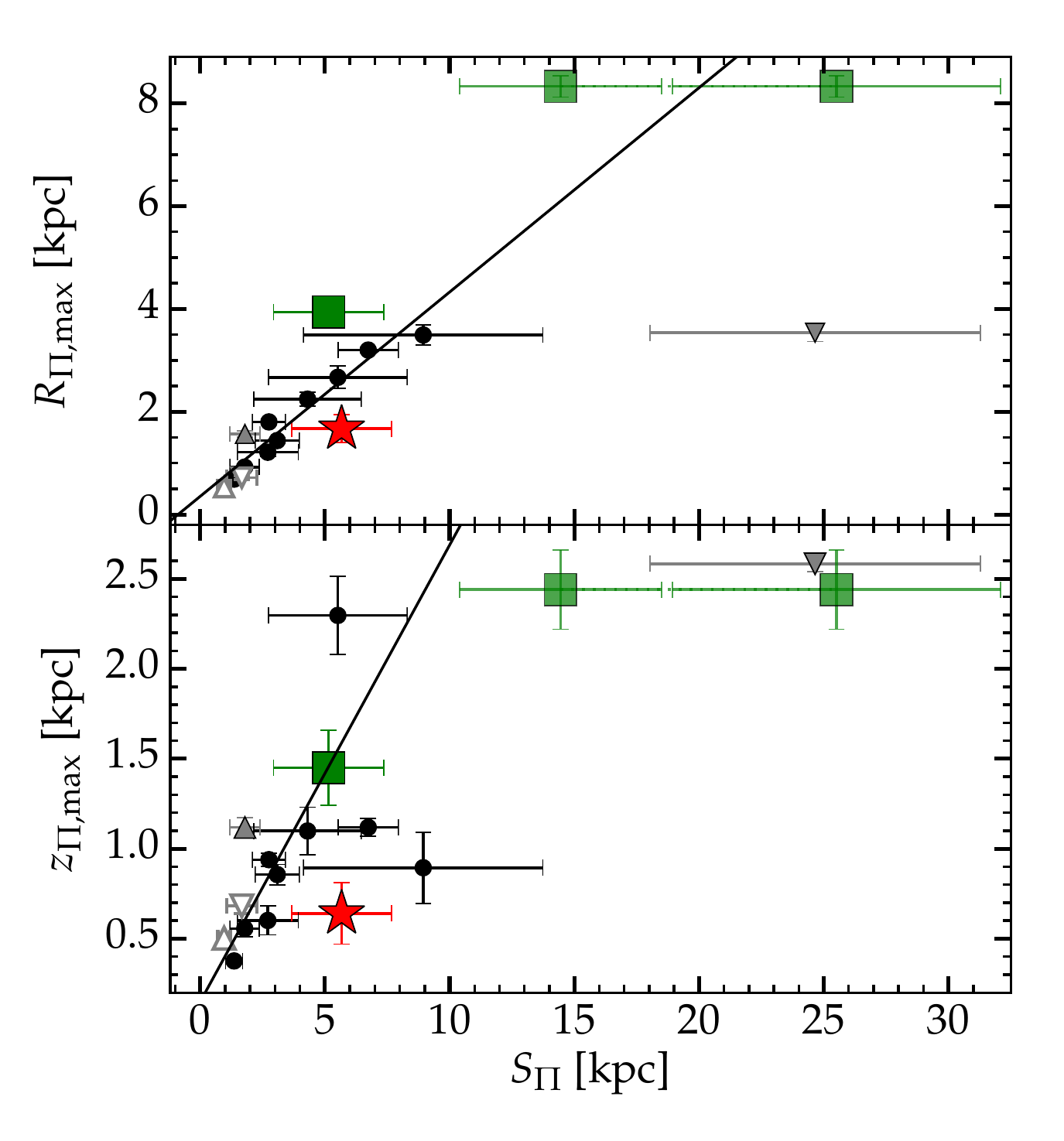} \includegraphics[width=.9\columnwidth]{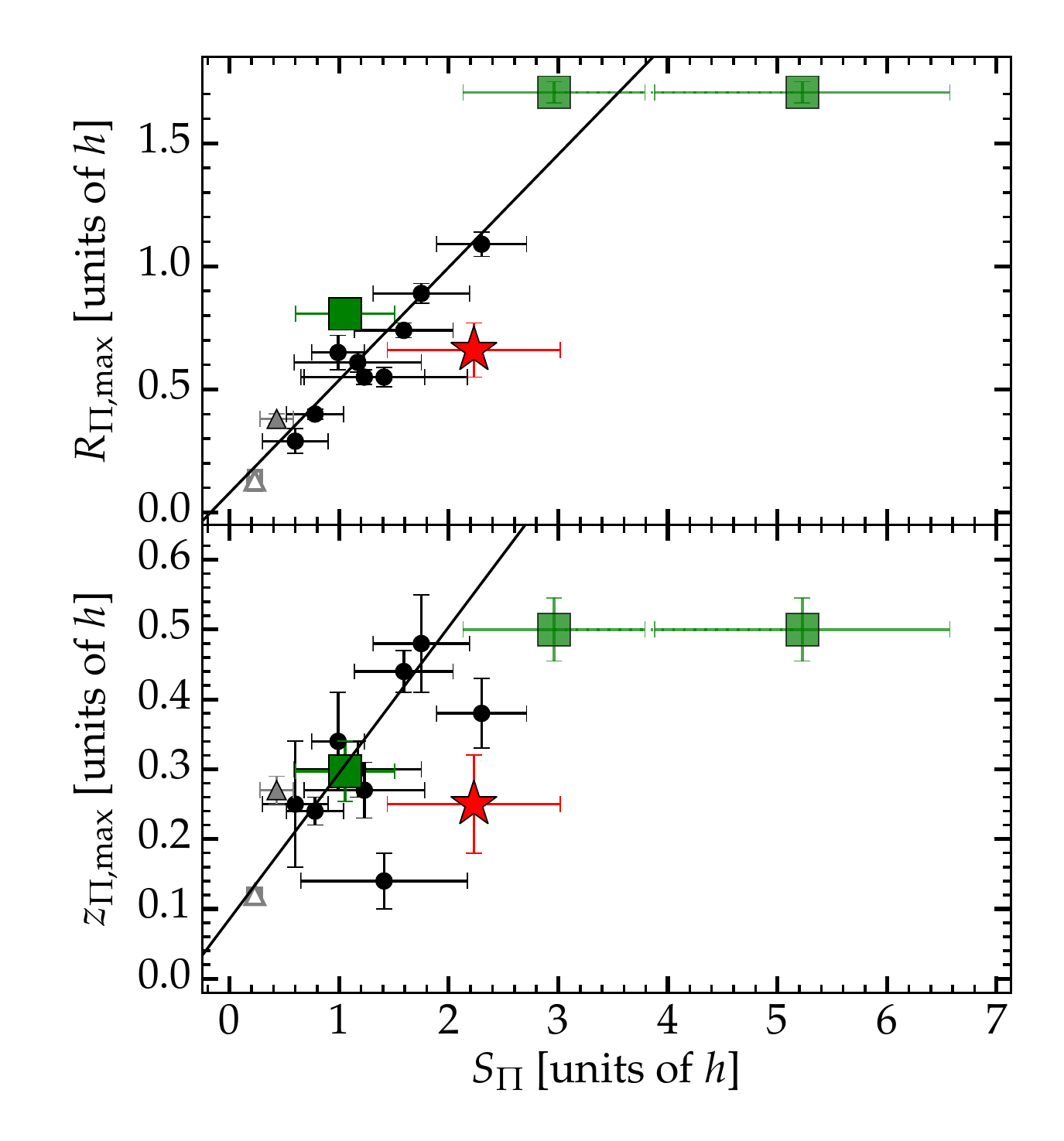}
\caption{CG16 scaling relations showing X/P radius (top) and height (bottom) as a function of integrated strength. The colour scheme is analogous to Figure \ref{fig:R_vs_z} and the lines represent linear fits from CG16. The correlations are shown in kpc (left) and in units of disc scale length $h$ (right). The outer peanut of NGC 128 is an outlier from the trends (outside the plotting area in the right-hand panels), possibly having its X/P strength enhanced through interactions with its satellite.}
\label{fig:zR_vs_S}
\end{figure*}

The size of X/P structures is also related to their integrated strength, as shown in Figure \ref{fig:zR_vs_S}. In order to exclude the effect of variation in galaxy size, this relation was additionally normalised by disc size (i.e., the exponential scale length $h$). The relation still holds, indicating the tight connection between X/P structures and the host discs they originate from. Here we employed the measured value of $h = 4.88$ kpc for full consistency with the observational approach. Once again it is apparent that the inner peanut in the simulation falls on the expected trend, whereas the outer X is significantly stronger than expected, in particular in the ($z_{{\rm\Pi,max}} - S_{\rm \Pi}$) plane. That is, it is a strong and longer than usual X-shaped structure, indeed extending all the way to the end of the bar, but which nonetheless remains relatively confined in the vertical direction. This is confirmed by the small opening angle of its X arms (Table \ref{tab:diagnostics}). We note that in Figure \ref{fig:zR_vs_S} the two data points corresponding to the outer X feature are the measurements on the thick and intermediate components, as the thin component was too noisy for a reliable measurement, and consequently the total image as well, as it is dominated by the latter component. As such, although strong, the two data points are lower limits of the total $S_{\rm \Pi}$ of the outer X structure. 

Comparing this simulation with nearby galaxies leads to conclude that the inner X/P feature is indeed a rather typical peanut, consistent with the theoretical formation scenarios, as well as with sizes (including fraction of the host bar size) and strengths of observed galaxies. Conversely, the outer structure is significantly stronger, sharper and has a X or ``bow-tie'', rather than peanut, morphology. Both its strength and size make it an outlier of the structural scaling relations defined by the CG16 sample, and the fact its maximal vertical extent edge-on corresponds roughly to the end of the bar (the scale of the face-on ansae) seems unusual. Nevertheless, this radial scale is also the location of the vILR, where a boxy/peanut feature is indeed expected to reach its maximal height (\citealt{Combes+1990}). Canonically, peanut structures are observed to arise as a thickening of roughly the inner half of bars ($\approx 0.4 - 0.5$ $R_{\rm bar}$; \citealt{LuettickeDettmar&Pohlen2000}, \citealt{Laurikainen&Salo2017}, \citealt{Erwin&Debattista2017}). It is therefore important to understand the formation and physical relevance of such extended X structures in $N$--body simulations which aim to explain and constrain the secular evolution of barred galaxies, as well as the Milky Way (e.g., \citealt{DiMatteo+2015}, \citealt{Debattista+2017}, \citealt{Fragkoudi+2017}, \citealt{DiMatteo+2019}). We reserve such an investigation for future works.

\section{Conclusions}\label{sec:conclusions}

In this paper we study a pure $N$--body simulation of a barred disc galaxy, consisting of a stellar disc embedded in a live dark matter halo, which develops two co-existing X/peanut-shaped structures, in a nested configuration, which form after $\approx$ 3.5 Gyr, and stabilise after $\approx$ 5 Gyr of evolution in isolation. The simulation initially consists of a pure disc made up of three kinematic components -- thin, intermediate, and thick. The absence of a classical bulge component\footnote{But not of a small $pseudo-$bulge, which indeed develops in the simulation.} facilitates the identification and quantification of the components of interest, particularly in the central region, where bulges otherwise dominate the light distribution in all but the later--type spiral galaxies. 

We extract and model the face--on major axis surface density profile from mock-images constructed from the simulation, analogously to a photometric decomposition of a galaxy image. We further extract the edge-on radial $B_6$ profile, which captures the signature of the two nested X/peanut--shaped structures, and use it to quantify their radial and vertical extents ($R_{\Pi}$, $z_{\Pi}$), as well as their respective integrated strength ($S_{\Pi}$). The main conclusions are as follows:

\begin{itemize}

\item The two X/P features are not associated with two distinct bars, but remain dynamically coupled throughout, rotating with the bar's pattern speed of $\Omega_p \approx 9$ km/s/kpc at $t=6.45$ Gyrs, a time where the configuration is in place, stable and symmetric. This is a rather slow bar, which only extends out to roughly half the radius of co-rotation. 

\item The inner, `classical' peanut--shaped structure develops following a vertical buckling instability of the bar. It peaks at $R = 3.94 \pm 0.21$ kpc, roughly half the radius of the bar, and its size and strength is consistent with those found in external galaxies.

\item The outer structure is considerably stronger, distinctly X-- or (bow-tie)--shaped, and peaks at $R = 8.33 \pm 0.22$ kpc, a scale roughly coinciding with the end of the bar and the scale of the latter's `handles', or ansae. While the vertical extent is not significantly higher than expected, the radius and strength make it an outlier of the trends seen in nearby galaxies. This feature develops shortly after the inner peanut, and gradually extends in radius, over the span of $\sim 1.5$ Gyr, from $\sim$ 5 kpc to its final stable radius just over 8 kpc, all the while showing little change in its vertical size.

\end{itemize}

We conclude that the two nested peanuts likely arise from inherently different dynamical mechanisms, appearing to be a superposition of the CX and OX morphological classes. Multiple X/P morphologies can be supported via various combinations of periodic, quasi-periodic, or chaotic orbits, as recently shown by \cite{Patsis&Harsoula2018} and \cite{Parul+2020}, and the absence of a classical bulge component to stabilise the slow bar in our simulation might facilitate the population of families that support a long-lasting double configuration. A detailed orbit frequency analysis is required to constrain the specific orbital backbones of the two structures, which however goes beyond the scope of this paper and we reserve for future work.

\section{Acknowledgements}

This work was supported by the Centre National d'\'Etudes Spatiales (CNES) post-doctoral funding project, and by the ANR (Agence Nationale de la Recherche) through the MOD4Gaia project (ANR-15- CE31-0007, PI: P. Di Matteo). This work was granted access to the HPC resources of CINES under the allocation A0020410154 made by GENCI. Funding for SDSS-III has been provided by the Alfred P. Sloan Foundation, the Participating Institutions, the National Science Foundation, and the U.S. Department of Energy Office of Science. Part of this work is based on observations made with the NASA/ESA $Hubble Space Telescope$, and obtained from the $Hubble Legacy Archive$, which is a collaboration between the Space Telescope Science Institute (STScI/NASA), the Space Telescope European Coordinating Facility (ST-ECF/ESA) and the Canadian Astronomy Data Centre (CADC/NRC/CSA).

%\clearpage

\bibliographystyle{mn2e}
\bibliography{C+20}

%\clearpage
%\appendix

%\setcounter{figure}{0}
%\renewcommand{\thefigure}{A\arabic{figure}}

%\section{Appendix}\label{sec:ap1} 
\end{document}